\documentclass{llncs}

\usepackage{psfrag}
\usepackage[vlined,boxed]{algorithm2e}

\usepackage{amsmath,amssymb,amstext,amsfonts}
\usepackage{graphicx}
\usepackage{pifont}

\newcommand{\comment}[1]{×}

\newcommand{\diss}[1]{}
\newcommand{\longversion}[1]{#1}

\newcommand{\dout}{d^+}
\newcommand{\din}{d^-}
\newcommand{\nout}{N^+}
\newcommand{\nin}{N^-}
\newcommand{\bvt}{\overline{V}_T}
\newcommand{\optprob}[3]{{\bf {\sc  #1}} \\
{\bf Given:} #2.\\
{\bf Task:} #3}
 \newcommand{\etal}{\emph{et al.}}
\def\rtdmlst{1.9043}

\newcommand{\NO}{\textup{\textsf{NO}}}

\SetAlgoSkip{}

\newcommand{\emphopt}[3]{{\bf {\sc  #1}} \\
{\bf Given:} #2.\\
{\bf Task:} #3}
\newcommand{\mc}{\mathcal}
\newcommand{\Oh}{{\mc{O}}}

\usepackage{tabularx}

\def\lab{\text{lab}}
\def\free{\text{free}}
\def\FL{\text{FL}}
\def\BN{\text{BN}}
\def\LN{\text{LN}}
\def\IN{\text{IN}}

\title{A Faster Exact Algorithm for the Directed Maximum Leaf Spanning Tree Problem} 

\author{Daniel Raible \&
      Henning Fernau
     }
\institute{%
Univ.
Trier, FB 4---Abteilung Informatik, 
54286 Trier,
Germany \\
\email{\{raible,fernau\}@informatik.uni-trier.de}
}

\begin{document}

\maketitle

\begin{abstract}
Given a directed graph $G=(V,A)$, the {\sc Directed Maximum Leaf Spanning Tree} problem asks to
compute a directed
 spanning tree (i.e., an out-branching) with as many leaves as possible. 
By designing a  Branch-and-Reduced algorithm combined with the
{\it Measure\&Conquer} technique for running time analysis, we show that the problem can be solved
in  time $\Oh^*(\rtdmlst^n)$ using polynomial space. Hitherto, there have been only few examples. Provided 
exponential space this run time upper bound can be lowered to $\Oh^*(1.8139^n)$.  
\end{abstract}

\diss{\chapter{A faster exact algorithm for the Directed Maximum Leaf Spanning Tree problem}} 

\section{Introduction}
We investigate the following problem
\emphopt{Directed Maximum Leaf Spanning Tree (DMLST)}{A directed graph  $G(V,A)$}{Find a directed spanning tree for $G$ with the maximum number of
leaves.}\\[1ex]
Alternatively, we can find an out-branching with  the maximum number of leaves. Here an out-branching in a directed graph is a spanning tree $T$ in
the
underlying undirected graph, but the arcs are directed from the root to the leaves, which are the vertices of out-degree zero with respect to $T$.
The terms out-branching and directed spanning tree are equivalent. 

\subsection{Known Results.}
The undirected version of the problem already has been widely studied with regard to its approximability. There is a 2-approximation running in polynomial
time by
R.~Solis-Oba~\cite{Sol-Oba98}. In almost linear time H.-I.~Lu and R.~Ravi~\cite{LuRav98} provide a 3-approximation.
P.S.~Bonsma and F.~Zickfeld~\cite{BonZic2008} could
show that the problem is $\frac{3}{2}$-approximable when the input is restricted to cubic graphs.
J.~Daligault and S.~Thomass\'e~\cite{DalTho2009} described a $92$-approximation algorithm together with 
an $\Oh(k^2)$-kernel for the {\sc Directed Maximum
Leaf Spanning Tree} problem. \\
This problem has also drawn notable attention in the field of parameterized algorithms. Here the problem is known as \emph{directed $k$-leaf
spanning tree} where $k$ is a lower bound on the number of leaves in the directed spanning tree.
The algorithm of J.~Kneis,
A.~Langer and P.~Rossmanith~\cite{KneLanRos2008a} solves this problem in time
 $\Oh^*(4^k)$\footnote{The notation $\Oh^*()$ suppresses polynomial factors.}.
Moreover, in J.~Daligault~\etal~\cite{DalGutKimYeo2008b}   an upper-bound of $\Oh^*(3.72^k)$ is achieved. 
The same authors could also analyze their algorithm with respect to the input size $n$. This implies a running time
 upper bound of $\Oh^*(1.9973^n)$.
D.~Raible and H.~Fernau~\cite{RaiFer2010} improved this running time to $\Oh^*(3.4575^k)$ in the more special 
case of undirected graphs.
\\
F.V.~Fomin, F.~Grandoni and D.~Kratsch~\cite{FomGraKra08} gave an exact, non-parameterized algorithm with 
run time $\Oh^*(1.9407^n)$ for the undirected
version. H.~Fernau \etal~\cite{Feretal2009b} improved this upper bound to $\Oh^*(1.8966^n)$.
I.~Koutis and R.~Williams~\cite{KouWil2009} could derive a randomized $\Oh^*(2^k)$-algorithm for the undirected version.
Using  an observation of V.~Raman and S.~Saurabh~\cite{RamSau2006} this implies a randomized algorithm with running time $\Oh^*(1.7088^n)$.

\subsection{Our Achievements.} The main result in this paper improves the current  best upper of$\Oh^*(1.9973^n)$ by \cite{DalGutKimYeo2008b}. We can
achieve a new bound of
$\Oh^*(\rtdmlst^n)$. Our algorithm is inspired by the one of \cite{Feretal2009b}. However, this algorithm cannot
be simply transferred to the directed
version. Starting from an initial root the algorithm grows a tree $T$. The branching process takes place by deciding whether the vertices neighbored
to the tree will become final leaves or internal vertices. A crucial ingredient of the algorithm was also to create \emph{floating leaves}, i.e.,
vertices which are final leaves in the future solution  but still have to be attached to the  $T$, the tree which is grown. This concept has been
already used
in \cite{Feretal2009b}  and partly by \cite{DalGutKimYeo2008b}. In the undirected case we guarantee that in the
bottleneck case we can generate at least two such leaves. In the directed version there is a situation where only one can be created.
Especially for this problem we had to find a workaround. 

\subsection{Preliminaries, Terminology \& Notation}
We consider directed graphs $G(V,A)$ in the course of our algorithm, where $V$ is the vertex set and $A$ the arc set. 
The \emph{in-neighborhood} of a vertex  $v \in V$ is $\nin_{V'}(v)=\{u \in V' \mid (u,v) \in A\}$ and, analogously, 
its \emph{out-neighborhood} is
$\nout_{V'}(v):=\{u \in V'\mid (v,u)\}$. The \emph{in- and out-degrees} of $v$ are $\din_{V'}(v):=|\nin_{V'}(v)|$ 
and $\dout_{V'}(v):=|\nout_{V'}(v)|$ and its
 \emph{degree} is $d_{V'}(v)=\din_{V'}(v)+\dout_{V'}(v)$. If $V'=V$ then we might suppress the subscript. For $V' \subseteq V$ 
we let $\nout(V):=\bigcup_{v
\in V'} \nout(v)$ and $\nin(V')$ is defined analogously.\\ 
Let $A(V'):=\{(u,v) \in A \mid \exists u,v \in V'\}$,
$\nout_A(v):=\{(v,u) \in A \mid u \in \nout_V(v)\}$ and $\nin_A(v):=\{(u,v) \in A \mid u \in \nin_V(v)\}$.
Given a graph $G=(V,A)$ and a graph $G'=(V',A')$, $G'$ is a \emph{subgraph}
of $G$ if $V' \subseteq V$ and $A' \subseteq A$.
The subgraph of $G$ induced by a vertex set $X \subseteq V$ is denoted by $G(X)$
and is defined by $G(X)=(X,A')$ where $A'=A(X)$.
The \emph{subgraph} of $G$ induced by an arc set $Y \subseteq A$ is denoted by $G(Y)$
and is defined by $G(Y)=(\tilde V,V(Y))$ where $V(Y)= \{ u\in V \mid \exists (u,v) \in Y \vee \exists (v,u) \in Y \}$.\\
A \emph{directed path} of length $\ell$ in $G$ is a set of pairwise different vertices $v_1,\ldots ,v_\ell$ such that $(v_i,v_{i+1}) \in A$ for $1 \le i < \ell$.
A subgraph $H(V_H,A_H)$ of $G$ is called a \emph{directed tree} if there is a unique root $r \in V_H$ such that there is a
unique directed path $P$  from $r$ to every $v \in V_H \setminus \{r\}$ under the restriction that its arc set obeys $A(P) \subseteq A_H$. Speaking
figuratively, in a
directed tree the arcs are directed from the parent to the child. If for a directed tree $H=(V_H,A_H)$ that is a subgraph of $G(V,A)$ we have $V=V_H$
we call it \emph{spanning
directed tree} of $G$. The terms \emph{out-tree} and \emph{out-branching} are sometimes used for directed tree and spanning
directed
tree, respectively. The \emph{leaves} of a directed tree $H=(V_H,A_H)$ are the vertices $u$ such that $\din_{V_H}(u)=d_{V_H}(u)=1$. In $leaves(H)$ all leaves
of a tree $H$ are comprised and $internal(H):=V(H) \setminus leaves(H)$.
The unique vertex $v$ such that $\nin_{V_H}(u)=\{v\}$ for a tree-vertex will be called \emph{parent} of $u$.
A vertex $v \in V_H$ such that $d_{V_H}(v) \ge 2$ will be called \emph{internal}. Let $T(V_T,A_T)$ and $T'(V_{T'},A_{T'})$ be two trees. $T'$
\emph{extends} $T$, written $T' \succeq T$, iff $V_T \subseteq V_{T'}$, $A_T \subseteq A_{T'}$. Simplistically, we will consider a tree $T$
also as a set of arcs $T \subseteq A$ such that $G(T)$ is a directed tree. The notions of $\succeq$ and $leaves(T)$ carry over canonically.\\
An \emph{arc-cut set} is a set of arcs $B \subset A$ such that $G(A \setminus B)$ is a digraph which is not connected. We
suppose that $|V|\ge 2$. The function $\chi()$ returns $1$ if its argument evaluates to true and $0$ otherwise.

\subsection{Basic Idea of the Algorithm}
First we formally re-define our problem:\\
\optprob{Rooted Directed Maximum Leaf Spanning Tree (RDMLST)}{A directed graph $G(V,A)$ and a vertex $r \in V$}{ Find a spanning directed tree $T'
\subseteq A$ such that $|leaves
(T')|$ is maximum and $\din_{T}(r)=0$.}\\[1ex]
Once we have an algorithm for  {\sc RDMLST} it is easy to see that it can be used to solve {\sc DMLST}.
As a initial step we will consider every vertex as a possible root $r$ of the final solution. This yields a total of $n$ cases.\\
Then in the course of the algorithm for {\sc RDMLST} we will gradually extend a out-tree $T \subseteq A$, which is predetermined to be a subgraph  in
the final out-branching.  Let $V_T:=V(T)$ and    $\bvt:= V \setminus V_T$. We will also maintain a mapping $lab: V \to
\{\free,\IN,\LN,\BN,\FL\}=:D$, which assigns different roles to the vertices. If
$lab(v)=\IN$ then $v$ is already fixed to be internal, if $lab(v)=  \LN$ then it will be a leaf. 
If $lab(v)=\BN$ then $v$ already has a parent in $T$, but can be leaf or internal in the final solution. In general we will decide this by branching
on such \BN-vertices. If $lab(v)=\FL$ then $v$ is constrained to be
a leaf but has not yet been attached to the tree $T$. Such vertices are called \emph{floating leaves}. If $lab(v)=\free$ then $v \not \in V_T$ and
nothing has been fixed or $v$ yet. For a \emph{label} $Z \in D$ and $v \in V$ we will often write $v \in
Z$ when we mean $lab(v)=Z$. Vertices in \IN\ or \LN\ will also be called \emph{internal nodes} or \emph{leaf nodes}, respectively. 
A given tree $T'$  defines a labeling $V_{T'} \to D$ to which we refer by $lab_{T'}$.. 
Let $\IN_{T'}:=\{v \in V_{T'} \mid \dout_{{T'}}(v) \ge 1\}$, $\LN_{T'}:=\{v \in V_{T'} \mid \dout_{{T'}}(v) = 0 \}$
 and $\BN_{T'}=V_{T'} \setminus (\IN_{T'} \cup \LN_{T'})$.
Then for any $ID \in D \setminus \{\FL,\free\}$ we have $ID_{T'}=lab^{-1}(ID)$.
 We always assure that $lab_T$ and $lab$ are the same on
$V_T$. 
The subscript might be hence suppressed if $T'=T$.
If $T' \succ T$, then we assume that $\IN_T \subseteq \IN_{T'}$ and $\LN_{T} \subseteq \LN_{T'}$. So,  the labels $\IN$ and $\LN$ remain once they are
fixed. 
For the remaining labels  we have the following possible
transitions: $\FL \to \LN$, $\BN \to \{\LN,\IN\}$ and $\free \to D \setminus \{\free\}$.
Let $\BN_i=\{v \in \BN \mid \dout(v)=i\}$, $\free_i=\{v \in \free \mid \din(v)=i \}$ for $i \ge 1$, $\BN_{\ge \ell}:= \cup_{j=\ell}^n \BN_j$ and
$\free_{\ge \ell}:= \cup_{j=\ell}^n \free_j$.
\section{The Polynomial Part}

\subsection{Halting Rules}
First we specify halting rules. If one of these rules applies the algorithm halts. Then it either returns a solution or answers that none can be
built in the according branch of the search tree.
\begin{description}
\item[(H1)] If there exists a $v \in \free \cup \FL$ with $d^-(v)=0$. Halt and answer \NO.
\item[(H2)] If $\BN = \emptyset$. Halt. A spanning tree has been constructed if $\free \cup \FL=\emptyset$. If so return $|\LN|$.

\item[(H3)] If there is a bridge $e:=(u,v) \in A \setminus T$ which splits the graph in at least two connected 
components of size at least two and $v \in \FL$. Halt and
answer \NO.
\end{description}

\subsection{Reduction rules}
We state a set of six reduction rules in the following. Similar reduction rules for the undirected version have already appeared in
\cite{Feretal2009b,RaiFer2010}. We assume that the halting rules are already checked exhaustively
\begin{description}
\item[(R1)] Let $v \in V$. If $lab(v)=\FL$ then remove $\nout_A(v)$. If $lab(v)=\BN$ then remove $\nin_A(v) \setminus T$.
\item[(R2)] If there exists a vertex $v \in \BN$ with $\dout(v)=0$ then set $\lab(v):=\LN$.
\item[(R3)] If there exists a vertex $v \in \free$ with $d(v)=1$ then set $\lab(v):=\FL$.
\item[(R4)] If $v \in \LN$ then remove $N_A(v) \setminus T$.
\item[(R5)]
Let $u \in \BN$ such that  $\nout_A(u)$ is a an arc-cut set. Then $\lab(u):=\IN$ and for all $x \in N^+(u) \cap \FL$ set $\lab(x):=\LN$,
and for all $x \in N^+(u) \cap \free$ set  $\lab(x):=\BN$.
\item[(R6)] If there is an arc $(a,b) \in A$ with $a,b \in \free$ and $G(A\setminus \{u,v\})$ consist of two strongly connected components of
vertex-size greater than
one. Then contract $(a,b)$. 
\end{description}
\begin{proposition}
 The reduction rules are sound.
\end{proposition}
\begin{proof}
 \begin{description}
\item[(R1)] A floating leaf $v$ cannot be a parent anymore. Thus, it is valid to remove $\nout_A(v)$. If $v \in \BN$ then $v$ already has a parent in
$T$. Thus, no arc in $\nin(v) \setminus T$ will ever be part of a tree $T' \succeq T$.
\item[(R2)] The vertex $v$ cannot be a parent anymore. Thus, setting $lab(v):=\LN$ is sound.
\item[(R3)] The vertex $v$ must be a leaf in any tree $T' \succeq T$.
\item[(R4)] The only arcs present in any tree $T' \succeq T$ will be $N_A(v) \cap T$. Thus, $N_A(v) \setminus T$ can be removed.
\item[(R5)] As $\nout_A(v)$ is an arc-cut set, setting $v \in \LN$ would cut off a component which cannot 
be reached from the root $r$. Thus, $v\in \IN$
is constrained.
  \item[(R6)]  Let $G^{*}$ be the graph after contracting $(h,u)$. If
$G^{*}$ has a spanning tree with $k$ leaves, then also $G$. On the other hand note that in every spanning tree $T' \succeq T$ for $G$   we have that
$h,u \in \IN$ and $(h,u) \in T'$. Hence, the tree $T^{\#}$ evolved by contracting $(h,u)$ in $T'$ is a spanning tree with $k$ leaves in $G^*$. 
 \end{description}

\end{proof}

\section{The Exponential Part}
\subsection{Branching rules}

If $\nout(internal(T )) \subseteq internal(T ) \cup leaves(T )$, we call $T$ an \emph{inner-maximal} directed tree. 
We make use of the following fact:
\begin{lemma}[\cite{KneLanRos2008a} Lemma 4.2] \label{drosslem1}
If there is a tree $T'$ with $leaves(T') \ge k$ such that $T' \succeq T$ and $x \in internal(T')$
 then there is a tree $T''$ with $leaves(T'') \ge
k$ such that $T'' \succeq T$, $x \in internal(T'')$ and $\{(x,u) \in A\} \subseteq T'' $ 
\end{lemma}

See the Algorithm~\ref{dalgo1} which describes the branching rules. As mentioned before, the search tree evolves by branching on \BN-vertices. For
some $v \in \BN$ we will set either $lab(v)=\LN$ or $lab(v)=\IN$. In the second case we adjoin the vertices $\nout_A(v) \setminus T$ as \BN-nodes to
the partial
spanning tree $T$. This is justified by Lemma~\ref{drosslem1}.  Thus, during the whole algorithm we only consider inner-maximal trees. Right in the
beginning we therefore  have $A(\{r\} \cup \nout(r))$ as a initial tree where $r$ is the vertex chosen as the root.

We also introduce an abbreviating notation for the different cases generated by branching: $\langle v \in \LN; v \in \IN \rangle $ means that we
recursively consider the two cases were $v$ becomes a leaf node and an internal node. The semicolon works as a delimiter between the different cases.
Of course, more complicated expression like $\langle v\in \BN, x \in \BN; v \in \IN, x \in \LN; v \in \LN \rangle$ are possible, which generalize
straight-forward.
\begin{algorithm}
\caption{An Algorithm for solving {\sc RDMLST}}\label{dalgo1}
{
\Indm
\Indp
\dontprintsemicolon
\KwData{A directed graph $G=(V,A)$ and a tree $T \subseteq A$.}
\KwResult{A spanning tree $T'$ with the maximum number of leaves such $T' \succeq T$}
Check if a halting rule applies.\\
Apply the reduction rules exhaustively.\\
	\If{$\BN_1 \neq \emptyset$}
       {
          Choose some $v \in \BN_1$.\\
	Let $P=\{v_0,v_1,\ldots ,v_k\}$ be a path of maximum length such that  $(1)$ $v_0=v$, for all $1 \le i \le k-1$
$(2)$ $\dout_{\overline{P_{i-1}}}(v_i)=1$ (where $P_{i-1}=\{v_0,\ldots
,v_{i-1}\}$) and $(3)$ $P \setminus \free \subseteq \{v_0,v_k\}$\\
	    
		\If{$\dout_{\overline{P_{i-1}}}(v_k)=0$}
		  {
		    Put $v \in \LN$ \hfill(B1)
		}
		\Else
		{
		$\langle v \in \IN, v_1,\dots, v_{k} \in \IN; v \in \LN \rangle$\hfill(B2)\;
		}
       }
	
	 \Else {Choose a vertex $v \in \BN$ with maximum out-degree.\\
	\If{{\it a)} $d^+(v) \geq 3$ \textbf{or} {\it b)}($\nout(v)=\{x_1,x_2\}$ and $\nout(v) \subseteq \FL$)}
	{
		$\langle v \in \IN; v \in \LN \rangle$ and in case {\it b)} apply {\tt makeleaves}($x_1,x_2$) in the 1st branch.\hfill(B3)\;
	}
	\ElseIf{$\nout(v)=\{x_1,x_2\}$}
	{ 
		  \If{for $z \in (\{x_1,x_2\} \cap \free)$ we have \\
		    $|N^+(z) \setminus  \nout(v)|=0$ \textbf{or}\hfill(B4.1)\\ 
		      $\nout_A(z)  \text{ is an arc-cut set}$ {\bf or} \hfill(B4.2)\\
			$N^+(z) \setminus  \nout(v)= \{v_1\}$. \hfill(B4.3)\\}
		  {
		    $\langle v \in \IN; v \in \LN \rangle$\hfill(B4)\;
		  }
	    }
	\ElseIf{$\nout(v) = \{x_1,x_2\}$, $x_1  \in \free, x_2 \in \FL$}
        {
          
	    $\langle v \in \IN, x_1 \in \IN; v \in \IN, x_1 \in \LN;v \in
		\LN \rangle$ and apply {\tt makeleaves}($x_1,x_2$) in the 2nd branch.\hfill(B5)\;

	  }
	\ElseIf{$\nout(v) = \{x_1,x_2\}$, $x_1,x_2 \in \free$,\\ $\exists z \in (\nin(x_1) \cap \nin(x_2))\setminus \{v\}$}
	      {
                        $\langle v \in \IN, x_1 \in \IN; v \in \IN, x_1 \in \LN,x_2 \in \IN;v \in
		\LN \rangle$\hfill(B6)\;
		}
	\ElseIf{$\nout(v) = \{x_1,x_2\}$, $x_1,x_2 \in \free$, $|(\nin(x_1) \cup \nin(x_2)) \setminus \{v,x_1,x_2\}|\ge 2$}
		{
				$\langle v \in \IN, x_1 \in \IN; v \in \IN, x_1 \in \LN, x_2 \in \IN; v\in \IN, x_1 \in \LN, x_2 \in \LN; v \in \LN
\rangle$ and apply {\tt makeleaves}($x_1,x_2$) in the 3rd branch.\hfill(B7)\;
			
		}
	      
	      \Else{ $\langle v \in \IN; v \in \LN \rangle$\hfill(B8)\;}

}
}
\end{algorithm}

\begin{procedure}
\Begin{$\forall u \in [(\nin(x_1) \cup \nin(x_2)) \setminus \{x_1,x_2,v\}] \cap \free$ set $u \in \FL;$\\
$\forall u \in [(\nin(x_1) \cup \nin(x_2)) \setminus \{x_1,x_2,v\}] \cap \BN$ set $u \in \LN;$
}
\caption{makeleaves($x_1,x_2$)}
\end{procedure}

\subsection{Correctness of the algorithm}

In the following we are going to prove a lemma which is crucial for the correctness and the running time.

\begin{lemma}\label{correct}
Let $T \subseteq A$ be a given tree such that $v \in \BN_T$ and $\nout(v)=\{x_1,x_2\}$. Let $T',T^\ast \subseteq A$ be  optimal solutions  with
$T',T^\ast \succeq T$ under the restriction that $lab_{T'}(v)=\LN$, and $lab_{T^\ast}(v)=\IN$ and $lab_{T^\ast}(x_1)=lab_{T^\ast}(x_2)=\LN$.

\begin{enumerate}
\item If there is a vertex $u \neq v$ with $\nout(u)=\{x_1,x_2\}$. Then $|leaves(T')|\ge |leaves(T^\ast)|$.\label{correct1} 
\item Assume that $\din(x_i) \ge 2$ ($i=1,2$). Assume that there exists some  $u \in (\nin(x_1) \cup \nin(x_2)) \setminus \{v,x_1,x_2\}$ such that 
$lab_{T^\ast}(u)= \IN$. Then
$|leaves(T')| \ge |leaves(T^\ast)|$.\label{lem2.2}\label{correct2}
\end{enumerate}

\end{lemma}

\begin{proof}
\begin{enumerate}
\item 
 Let $T^+:=(T^\ast \setminus \{(v,x_1),(v,x_2)\}) \cup \{(u,x_1),(u,x_2)\}$. We have $lab_{T^+}(v)=\LN$ and $u$ is the only vertex besides $v$ where
$lab_{T^\ast}(u) \neq lab_{T^+}(u)$ is possible. Hence, $u$ is the only vertex where we could have $lab_{T^\ast}(u)=\LN$ such that $lab_{T^+}(u)=
\IN$. Thus, we can conclude $|leaves(T^+)|\ge |leaves(T^\ast)|$. As $T'$ is optimal under the restriction that $v \in \LN$ it
follows
$|leaves(T')| \ge
|leaves(T^+)| \ge  |leaves(T^\ast)|$. 
\item W.l.o.g. we have $u \in \nin(x_1)\setminus \{v,x_2\}$. Let $q \in \nin(x_2) \setminus \{v\}$ and $T^+:=(T^\ast \setminus \{(v,x_1),(v,x_2)\})
\cup
\{(u,x_1),(q,x_2)\}$. We have $lab_{T^+}(v)=\LN$,
$lab_{T^+}(u)=lab_{T^\ast}(u)= \IN$ and $q$ is the only vertex besides $v$ where we could have $lab_{T^\ast}(q)\neq lab_{T^+}(q)$ (i.e.,
$lab_{T^\ast}(q)=\LN$
and $lab_{T^+}(q)=\IN$). Therefore $|leaves(T')| \ge
|leaves(T^+)| \ge |leaves(T^\ast)|$.
\end{enumerate}\qed
\end{proof}

\subsubsection{Correctness of the Different Branching Cases}
First note that {\bf (H2)} takes care of the case it indeed an out-branching has been built. If so the number of its leaves is returned.

Below we will argue that each branching case in Algorithm~\ref{dalgo1} is correct in a way  that it
preserves at least one optimal solution. Cases  (B4) and (B8) do not have to be considered in detail as these 
are simple binary and exhaustive branchings.
\begin{description}
\item[(B1)] Suppose there is an optimal extension $T' \succeq T$ such that $lab_{T'}(v)=lab_{T'}(v_0)=\IN$. Due to the structure of $P$ there must be
an $i$, $0<i \le k$ such that $(v_j,v_{j-1}) \in T'$ for $0<j \le i$, i.e., $v,v_1,\ldots v_{i-1} \in \IN$ and $v_i \in \LN$. W.l.o.g., we choose
$T'$ in a way that $i$ is minimum but $T'$ is still optimal (\ding{59}). By {\bf (R5)} there must be a vertex $v_z$, $0<z \le i$, such that there is an arc $(q,v_z)$ with $q
\not
\in P$. Now consider $T''=(T' \setminus \{(v_{z-1},v_z)\}) \cup \{q,v_z\}$. In $T''$ the vertex  $v_{z-1}$ is a leaf and therefore $|leaves(T'')|  \ge
|leaves(T')|$.
Additionally, we have that $z-1 <i$ which is a contradiction to the choice of $T'$ (\ding{59}).
\item[(B2)] Note that $lab(v_{k})\in \{\BN,\FL \}$ is not possible due to {\bf (R1)} and, thus, $lab(v_{k})=\free$. By the above arguments from {\bf (B1)} we can
exclude the case  that $v,v_1,\ldots v_{i-1} \in \IN$ and $v_i \in \LN$ ($i \le k$). Thus, under the restriction
that we set $v \in \IN$, the only
remaining possibility is also to set $v_1,\ldots v_{k} \in \IN$.
\item[(B3)] {\it b)} When we set $v \in \IN$ then the two vertices in $\nout(v)$ will become leaf nodes (i.e., become part of \LN). Thus,
Lemma~\ref{correct}.\ref{correct2} applies (Note that {\bf (R5)} does not apply and therefore $(\nout(x_1) \cup \nout(x_2))\setminus
\{v,x_1,x_2\}\neq \emptyset$).
This means that that every
vertex in $(\nin(x_1) \cup \nin(x_2)) \setminus \{v,x_1,x_2\}$ can
be assumed a to be leaf node in the final solution. This justifies to apply {\tt makeleaves}($x_1,x_2$).
\item[(B5)] The branching is exhaustively with respect to $v$ and $x_1$. Nevertheless, in the second branch {\tt makeleaves}($x_1,x_2$) is carried
out. This is justified by Lemma~\ref{correct}.\ref{correct2} as by setting $v \in \IN$ and $x_1 \in \LN$, $x_2$ will
be attached to $v$ as a \LN-node and {\bf (R5) } does not apply.
\item[(B6)] In this case we neglect the possibility that $v \in \IN, x_1,x_2 \in \LN$. But due to Lemma~\ref{correct}.\ref{correct1} a no worse
solution can be found in the recursively considered case  where we set $v \in \LN$. This shows that the  considered cases are sufficient.
\item[(B7)]Similarly, as in case (B3) we can justify  by Lemma~\ref{correct}.\ref{correct2}   the application of {\tt makeleaves}($x_1,x_2$) in the
third branch.
\end{description}
Further branching cases will not be considered as their correctness is clear due to exhaustive  branching.

\subsection{Analysis of the Running Time }
\subsubsection{The Measure}
To analyze the running-time we follow the {\it Measure\&Conquer}-approach (see \cite{FomGraKra2009}) 
and use the following  measure:
$$\mu(G) =
\sum_{i=1}^n\epsilon_i^{\BN} |\BN_i| + \sum_{i=1}^n \epsilon_i^\free |\free_i|+
\epsilon^\FL|\FL|$$

The concrete values are $\epsilon^\FL=0.2251$, $\epsilon_1^\BN=0.6668$, $\epsilon_i^\BN=0.7749$ for $i \ge 2$, $\epsilon_1^\free=0.9762$  and
$\epsilon_2^\free=0.9935$. Also let $\epsilon_j^\free=1$ for $j \ge 3$ and\\
$\eta=\min\{\epsilon^\FL,(1-\epsilon_1^\BN),(1-\epsilon_2^\BN),(\epsilon_2^\free-\epsilon_1^\BN),(\epsilon_2^\free-\epsilon_2^\BN),
(\epsilon_1^\free-\epsilon_1^\BN),(\epsilon_1^\free-\epsilon_2^\BN)\}=\epsilon_1^\free - \epsilon_2^\BN=0.2013$.

For $i \ge 2$ let $\Delta_i^\free= \epsilon_{i}^\free-\epsilon_{i-1}^\free$ and $\Delta_1^\free=\epsilon_1^\free$. Thus, $\Delta_{i+1}^\free \le
\Delta_i^\free$ with $\Delta_s^\free=0$ for $s \ge 4$. 
\subsubsection{Run Time Analysis of the Different Branching Cases}\ \\
In the following  we state for every branching case by how much $\mu$ will be reduced. Especially,  $\Delta_i$ states the amount by which the $i$-th
branch decreases $\mu$. If $v$ is the vertex chosen by Algorithm~\ref{dalgo1} the it is true 
that for all $x \in \nout(v)$ we have $\din(x) \ge 2$ by {\bf (R5)} (\ding{60}). 

\begin{description}
 \item[(B2)] $\langle v \in \IN, v_1,\dots, v_k \in \IN,  v \in \LN \rangle$

	Recall that $d^+_{\overline{P_{k-1}}}(v_{k})\geq 2$ and  $v_{k} \in \free$ by {\bf (R1)}. Then we must have that $v_{1} \in\free_{\ge 2}$ by
{\bf(R5)}.

	 \begin{enumerate}   
	\item $v$ becomes $\IN$-node;
	$v_1, \dots, v_{k}$ become $\IN$-nodes;
	the free vertices in $\nout(v_{k})$ become $\BN$-nodes, the floating leaves in $\nout(v_{k})$ become \LN-nodes: \\
	$\Delta_1 =
	\epsilon_{1}^{\BN}+\sum_{i=2}^{k} \epsilon_1^\free
	+ \chi(v_{1} \in \free_2)\cdot \epsilon_2^\free+\chi(v_{1} \in \free_{\ge 3})\cdot \epsilon_3^\free +2\cdot \eta$

	\item $v$ becomes $\LN$-node; the degree of $v_1$ is reduced:\\
	$\Delta_2 = \epsilon_{1}^{\BN}+\sum_{i=2}^3 \chi(v_1 \in \free_i)\cdot \Delta_i^\free$
	\end{enumerate}
\item[(B3)]  $\langle v \in \IN; v \in \LN \rangle$.
	
	Case {\it a)}
	\begin{enumerate}
	\item $v$ becomes $\IN$-node;
	the $\free$ out-neighbors of $v$ become $\BN$-nodes;
	the $\FL$ out-neighbors of $v$ becomes $\LN$-nodes:
	
	$\Delta_1 =
	\epsilon_{2}^{\BN}
	+ \sum_{x \in \nout(v) \cap \free_{\ge 3}} (1-\epsilon_{2}^{\BN})+\sum_{x \in \nout(v) \cap \free_{2}}
(\epsilon_2^\free-\epsilon_{2}^{\BN})
	+ \sum_{y \in \nout(v) \cap \FL} \epsilon^{\FL}$\\[-1ex]
	
	\item $v$ becomes $\LN$-node; the in-degree of the free out-neighbors of $v$ is decreased;
	$\Delta_2 =
	\epsilon_{2}^{\BN}+\sum_{i=2}^3 |\nout(v) \cap \free_{i}|\cdot  \Delta^\free_i$
	
	\end{enumerate}
Case {\it b)}\\
Recall that $v$ is a $\BN$ of maximum out-degree, thus $\dout(z)\leq \dout(v)=2$ for all $z \in \BN$. On the other hand $\BN_1=\emptyset$
which implies $\BN=\BN_2$ from this point on. Hence, we have $\nout(v)=\{x_1,x_2\}$, $\din(x_i) \ge 2$, $(i=1,2)$ 
and $|(\nin(x_1) \cup
\nin(x_2))\setminus \{v,x_1,x_2\}|\ge 1$ by  (\ding{60}), in the
following branching cases.
Therefore the additional amount of $min
\{\epsilon_1^\free-\epsilon^{\FL},\epsilon_2^{\BN}\}$ in the first branch is justified by the application of 
{\tt makeleaves}$(x_1,x_2)$. Note that
by (\ding{60}) at least one \free-node becomes a \FL-node, or one \BN-node becomes a \LN-node.
Also due to
{\bf (R1)} we
have that $\nout(x_i)\cap \BN =\emptyset$. 

\begin{enumerate}
\item $v$ becomes \IN-node; the \FL\ out-neighbors of $v$ become \LN-nodes; the vertices in $[\nin(x_1) \cup \nin(x_2) \setminus \{v,x_1,x_2\}] \cap
\BN$
become \LN-nodes; the vertices in $[\nin(x_1) \cup \nin(x_2) \setminus \{v,x_1,x_2\}] \cap \free $
become \FL-nodes.\\ $\Delta_1=\epsilon_2^\BN+2\cdot \epsilon^\FL + min \{\epsilon_1^\free-\epsilon^{\FL},\epsilon_2^{\BN}\}$
\item $v$ becomes \LN; $\Delta_2=\epsilon_2^\BN$.\\[1ex]
\end{enumerate}

\item[(B4)]$\langle v \in \IN; v \in \LN \rangle$.
\begin{enumerate}
\item[(B4.1):] 
\begin{enumerate} 
\item[1.] $v$ becomes \IN-node; $z$ becomes \LN-node by {\bf (R1)}, {\bf (R2)} or both {\bf R4)}; The vertex $q \in \{x_1,x_2\} \setminus\{z\}$
becomes
\LN-node or \BN-node (depending on $q \in \FL$ or $q
\in \free$)\\
 $\Delta_1=\epsilon_2^{\BN}+\epsilon_2^\free+\min\{\epsilon^\FL,(\epsilon_2^\free-\epsilon_2^{\BN})\}$ 
\item[2.]  $z$ becomes \LN-node;\\
$\Delta_2=\epsilon_2^\BN$ 
\end{enumerate}
\item[(B4.2):] 
\begin{enumerate} 
\item[1.] $v$ becomes \IN-node; $(z,h)$  $\nout_A(z)$ is an arc-cut. Thus, $z$ becomes \IN-node as {\bf (R5)} applies; The vertex $q \in
\{x_1,x_2\} \setminus\{z\}$ becomes \LN-node
or \BN-node
(depending on $q \in \FL$ or $q \in \free$)\\
 $\Delta_1=\epsilon_2^{\BN}+\epsilon_2^\free+\min\{\epsilon^\FL,(\epsilon_2^\free-\epsilon_2^{\BN})\}$ 
\item[2.]  $z$ becomes \L
N-node;\\
$\Delta_2=\epsilon_2^\BN$
\end{enumerate}
Note that in all following branching cases we have $\nout(x_i) \cap \free_1 = \emptyset$ ($i=1,2$) by this case.

\item[(B4.3):]
We have $|N^+(z) \setminus \nout(v)|= 1$. Thus, in the next recursive call after the first branch and the exhaustive
 application of {\bf (R1)}, either {\bf (R6)}, case (B2) or (B1) applies.
{\bf (R5)} does not apply due to (B4.2) being ranked higher.
 Note that the application of any other reduction rule does not change the situation. If (B2) applies
we can analyze
the current case  together with its succeeding one. If (B2) applies in the case we set $v \in \IN$ we deduce that $v_0,v_1,\ldots,v_{k} \in
free$ where $z=v_0=x_1$ (w.l.o.g., we assumed $z=x_1$). Observe that $v_1 \in \free_{\ge 2}$ as (B4.2) does not apply.
\begin{enumerate}
\item[1.] $v$ becomes \IN-node; $x_1$ becomes  \LN-node; $x_2$ becomes \FL- or \BN-node (depending on whether $x_2 \in \free$ or $x_2 \in \FL$; the
degree of $v_1$
drops:
\\[1ex] $\Delta_{11}=\epsilon_2^\BN+   \chi(x_1 \in \free_{\ge 3})\cdot \epsilon_3^\free + \chi(x_1 \in
\free_2)\cdot \epsilon_2^\free  + \\ 
\chi(x_2 \in \free_{\ge 3})\cdot (\epsilon_3^\free -\epsilon_2^\BN)+ \chi(x_2 \in \free_2)\cdot
(\epsilon_2^\free -\epsilon_2^\BN) + \chi(x_2 \in \FL)\cdot\epsilon^\FL + \sum_{i=2}^3 \chi(v_1 \in \free_i)\cdot \Delta_i^\free$
\item[2.] $v$ becomes \IN-node, $x_1,v_1 \in \IN, \ldots , v_{k}$ become \IN-nodes;  the free vertices in $\nout(v_{k})$ become $\BN$-nodes, the
floating leaves in
$\nout(v_{k})$ become \LN-nodes:\\[1ex]
$\Delta_{12}=\epsilon_2^\BN+\chi(x_1 \in \free_{\ge 3})\cdot
\epsilon_3^\free+\chi(x_1 \in \free_2)\cdot \epsilon_2^\free+\\
\chi(x_2 \in \free_{\ge 3})\cdot (\epsilon_3^\free -\epsilon_2^\BN)+ \chi(x_2 \in \free_2)\cdot
(\epsilon_2^\free -\epsilon_2^\BN) + \chi(x_2 \in \FL)\cdot\epsilon^\FL+\\
\chi(v_1 \in \free_2)\cdot \epsilon_2^\free+\chi(v_1 \in \free_{\ge 3})\cdot
\epsilon_3^\free+\sum_{i=2}^{k} \epsilon_1^\free + 2 \eta$
\item[3.] $v$ becomes \LN-node: the degrees of $x_1$ and $x_2$  drop:\\  
 $\Delta_2=\epsilon_2^\BN + \sum_{\ell=2}^{\max\limits_{h\in \{1,2\}}\din(x_h)}\sum_{j=1}^2  \chi(x_j \in \free_\ell)\cdot \Delta_\ell^\free$.
\end{enumerate}

If case (B1) applies to $v_1$ the reduction in both branches is as least as great as in (B4.1)/(B4.2). \\
If {\bf (R6)} applies after the first branch (somewhere in the graph) we get  
$\Delta_1=\epsilon_2^{\BN}+(\epsilon^\free_2-\epsilon_1^\BN)+\epsilon_1^\free+\min\{\epsilon^\FL,(\epsilon_2^\free-\epsilon_2^{\BN})\}$ and
$\Delta_2=\epsilon_2^\BN$. Here the amount of $\epsilon_1^\free$ in $\Delta_1$ originates from an {\bf (R6)} application.
\end{enumerate}

\item[(B5)] $\langle v \in \IN, x_1 \in \IN; v \in \IN, x_1 \in \LN;v \in
		\LN \rangle$
\begin{enumerate}
\item $v$ and $x_1$  become \IN-nodes; $x_2$ becomes a \FL-node; the vertices in $\nout(x_1) \cap \free$ become \BN-nodes; the vertices in $\nout(x_1)
\cap \FL$
become \LN-nodes;\\[1ex]
$\Delta_1 =\epsilon_2^{\BN}+\epsilon_2^\free + \epsilon^\FL +\sum_{x \in \nout(x_1) \cap \free} (\epsilon_2^\free-\epsilon_{2}^{\BN})+
      \sum_{x \in \nout(x_1) \cap \FL} \epsilon^{\FL}$\\

\item $v$ becomes \IN-node; $x_1$ becomes \LN-node; $x_2$ becomes \LN-node; after applying {\tt makeleaves}$(x_1,x_2)$ the vertices in $[\nin(x_1)
\cup \nin(x_2) \setminus \{v,x_1,x_2\}] \cap \BN$ become \LN-nodes and the vertices in $[\nin(x_1) \cup \nin(x_2) \setminus \{v,x_1,x_2\}] \cap \free
$
become \FL-nodes:\\[1ex]
$\Delta_2=\epsilon_2^{\BN} +\epsilon_2^\free  + \epsilon^\FL+ \min \{\epsilon_1^\free-\epsilon^{\FL},\epsilon_2^{\BN}\}$
\item $v$ becomes \LN: $\Delta_3=\epsilon_2^{\BN}$
\end{enumerate}
The amount of $\min\{\epsilon^\FL,(\epsilon_1^\free-\epsilon_2^\BN)\}$ in the second branch is due to (\ding{60}).

\item[(B6)] $\langle v \in \IN, x_1 \in \IN; v \in \IN, x_1 \in \LN, x_2 \in \IN;v \in \LN \rangle$ The branching vector can be derived by considering
 items 1,2
and 4 of (B7) and the reductions $\Delta_1,\Delta_2$ and $\Delta_4$ in $\mu$ obtained in each item.

\item[(B7)]  $\langle v \in \IN, x_1 \in \IN; v \in \IN, x_1 \in \LN, x_2 \in \IN; v\in IN, x_1 \in \LN, x_2 \in \LN; v \in \LN \rangle$

	Note that if $\nout_A(x_1)$ or $\nout_A(x_2)$ is an arc-cut set  then (B4.2) applies. Thus, all the
branching cases must be applicable. 
	
	Moreover due to the previous branching case (B4.3) we have $|\nout(x_1) \setminus \nout(v)|=|\nout(x_1) \setminus \{x_2\}|\ge 2$ and
$|\nout(x_2)
\setminus \nout(v)|=|\nout(x_2) \setminus \{x_1\}|\ge 2$ (\ding{81}).

Note that $\nin(x_1) \cap \nin(x_2)=\{v\}$ due to (B6).
\\[1ex] For $i \in \{1,2\}$ let $fl_i = |\{x \in \nout(x_i) \setminus \nout(v)\mid x \in FL\}|$, $fr_i^{\ge 3}=|\{u \in \nout(x_i) \setminus
\nout(v)\mid u \in \free_{\ge 3}\}|$ and
$fr_i^{2}=|\{u \in \nout(x_i) \setminus \nout(v)\mid u \in \free_{2}\}|$. \\
Observe that for $i \in \{1,2\}$ we have $(fl_i+fr_i^{\ge 3} +fr_i^2) \ge 2$ due to (\ding{81}).\\
	\begin{enumerate}
	\item $v$ becomes $\IN$;
	$x_1$ becomes $\IN$;
	$x_2$ becomes $\BN$;
	the $\free$ out-neighbors of $x_1$ become $\BN$;
	the $\FL$ out-neighbors of $x_1$ become $\LN$;
	
	$\Delta_1 =
	\epsilon_{2}^{\BN}
	+ \chi(x_1 \in \free_{\ge 3})+\chi(x_1 \in \free_2)\cdot \epsilon_2^\free
	+ \chi(x_2 \in \free_{\ge 3})\cdot (\epsilon_3^\free-\epsilon_{2}^{\BN})+ \chi(x_2 \in \free_{2})\cdot (\epsilon_2^\free-\epsilon_{2}^{\BN})\\
	+ (fl_1\cdot \epsilon^\FL + fr_1^{\ge 3}\cdot (\epsilon_3^\free-\epsilon_2^\BN)+ fr_1^{2}\cdot (\epsilon_2^\free-\epsilon_2^\BN))$
	\item $v$ becomes $\IN$;
	$x_1$ becomes $\LN$;
	$x_2$ becomes $\IN$;
	the $\free$ out-neighbors of $x_2$ becomes $\BN$;
	the $\FL$ out-neighbors of $x_2$ become $\LN$;

	$\Delta_2 =
	\epsilon_{2}^{\BN}+
	(\sum_{i=1}^2 [\chi(x_i \in \free_{\ge 3})\cdot \epsilon_3^\free+\chi(x_i \in \free_2)\cdot \epsilon_2^\free]) \\
	+ (fl_2\cdot \epsilon^\FL + fr_2^{\ge 3}\cdot (\epsilon_3^\free-\epsilon_2^\BN)+ fr_2^{2}\cdot (\epsilon_2^\free-\epsilon_2^\BN))$
		
	\item $v$ becomes $\IN$;
	$x_1$ becomes $\LN$;
	$x_2$ becomes $\LN$;
	the $\free$ in-neighbors of $x_1$ become $\FL$;
	the $\BN$ in-neighbors of $x_1$ become $\LN$;
	the $\free$ in-neighbors of $x_2$ become $\FL$;
	the $\BN$ in-neighbors of $x_2$ become $\LN$:\\

	$\Delta_3 =
	\epsilon_{2}^{\BN}+
	[\sum_{i=1}^2 (\chi(x_i \in \free_{\ge 3})\cdot \epsilon_3^\free+\chi(x_i \in \free_2)\cdot \epsilon_2^\free)] \\
	+ \max\{2,(\din(x_1)+\din(x_2)-4)\} \cdot \min \{\epsilon_1^\free-\epsilon^{\FL},\epsilon_2^{\BN}\}$\\[1ex]
	Note that the additional amount of  $\max\{2,(\din(x_1)+\din(x_2)-4)\}\cdot \{\epsilon_2^\free-\epsilon^{\FL},\epsilon_2^{\BN}\}$ is justified
by Lemma~\ref{correct}.\ref{correct2} and by the fact that $\din(x_i)\ge 2$ and $\nin(x_1) \cap \nin(x_2) =\{v\}$ due to (B6). Thus, we have 
$|\nin(x_1) \cup \nin(x_2)\setminus \{x_1,x_2,v\}|\ge \max\{2,(\din(x_1)+\din(x_2)-4)\}$. 
	\item $v$ becomes $\LN$; the degrees of $x_1$ and $x_2$ drop:\\
	$\Delta_4 =
	\epsilon_{2}^{\BN}+ \sum_{j=2}^{\max\limits_{\ell \in \{1,2\}}\{\din(x_\ell)\}} \sum_{i=1}^2 (\chi(\din(x_i)=j)\cdot \Delta_{j}^\free  ) 
$	
	\end{enumerate}

\item[(B8)]
 Observe that in the second branch we can apply {\bf (R6)}. Due to the non-applicability of {\bf (R5)} and the fact that (B7) is ranked
higher in priority we  have $|(\nin(x_1) \cup \nin(x_2)) \setminus
\{v,x_1,x_2\}|=1$. Especially,  (B6) cannot be applied by which we derive that $\nin(x_1) \cap \nin(x_2)=\{v\}$. Thus, due to this we have the
situation in Figure~\ref{problem}. 

So, w.l.o.g, there are  arcs $(q,x_1),(x_1,x_2)\in A$, where $\{q\}=(\nin(x_1) \cup \nin(x_2)) \setminus
\{v,x_1,x_2\}$, because we can rely on $\din(x_i)\ge 2$ ($i=1,2$) by (\ding{60}).
\begin{enumerate}
 \item  Firstly, assume that $q \in \free$. 
\begin{enumerate}
\item  $v$ becomes \IN; $x_1$ and $x_2$ becomes \BN:\\[1ex]
$\Delta_1=\epsilon_2^{\BN}+2 \cdot
(\epsilon_2^\free-\epsilon_2^{\BN})$
\item The arc $(q,x_1)$ will be contracted by {\bf (R6)} when we  $v$ becomes   \LN, as $x_1$ and $x_2$ only
can be reached by using $(q,x_1)$:\\[1ex]
 $\Delta_2=\epsilon^{\BN}_2+\epsilon_1^\free$.
\end{enumerate}
\item Secondly, assume $q \in \BN$. Then $q  \in \BN_2$ due to the branching priorities.
\begin{enumerate}
\item $v$ becomes \IN; $x_1$ and $x_2$ become \BN:\\[1ex]
$\Delta_1=\epsilon_2^{\BN}+2 \cdot
(\epsilon_2^\free-\epsilon_2^{\BN})$
\item 
 Then after setting $v \in\LN$, rule {\bf (R5)} will make $q$
internal  and subsequently also $x_1$:\\[1ex]
$\Delta_2=\epsilon_2^\BN+\epsilon_2^\free+\epsilon_2^\BN$.\\
This amount is justified by the changing roles of the vertices in $\nout(q) \cup \{q\}$.
\end{enumerate}

\end{enumerate}

\end{description}
\begin{figure}[ht]
\psfrag{q}{$q$}
\psfrag{x1}{$x_1$}
\psfrag{x2}{$x_2$}
\psfrag{v}{$v$}
\centering
\includegraphics[scale=0.75]{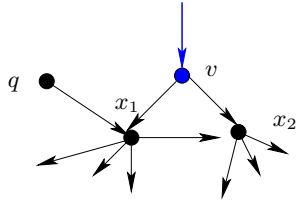}
\caption{The only situation which can occur in branching case (B8). The blue arc is contained in $T$.}
\label{problem}
\end{figure}
By the above case analysis we are able to conclude:
\begin{theorem}
 {\sc Directed Maximum Leaf Spanning Tree} can be solved in $\Oh^*(\rtdmlst^n)$ steps.
\end{theorem}
The proven run time bound admits only a small gap to the bound of $\Oh^*(1.8966^n)$ for the undirected version.
It seems that we can benefit from degree two vertices only on a small scale in contrast to the undirected problem version. Speaking loosely if $v
\in \BN_2$ and $x \in N(v)$ we can follow a WIN/WIN
approach in the undirected version. Either $d(x)$ is quite big then we will add many vertices to \BN\ or \FL\ when $v$ and subsequently $x$ become
internal. If $d(x)$ is
small, say two, then by setting $v \in \LN$ the vertex $x$ becomes a \FL-node. This implies also an extra reduction of the measure. We point out that
in the directed case the in- and out-degree of a
vertex generally is not related. Thus, the approach described for the undirected problem remains barred for the
 directed version. 

\section{Conclusions}
\subsection{An Approach Using Exponential Space}
The algorithm of J.~Kneis \etal~\cite{KneLanRos2008a} can also be read in an exact non-parameterized way. It is not hard to see that it yields a running
time of $\Oh^*(2^n)$. Alternatively, keep the cases (B1) and (B2) of Algorithm~\ref{dalgo1} and substitute all following cases by a simple
branch on
some \BN-node. Using $n$ as a measure we see that $\Oh^*(2^n)$ is an upper bound. \\
We are going to use the technique of memoization to obtain an improved running time. Let $SG^\alpha:=\{G(V') \mid V' \subseteq V, |V'| \le \alpha
\cdot
n\}$ where $\alpha=0.141$. Then we aim to create the following table $L$ indexed by some $G' \in SG^\alpha$ and some $V_{\BN} \subseteq V(G')$:
\\[1ex]
$
L[G',V_\BN]=T'\text{ such that } |leaves(T')|=\min_{\tilde T \subseteq {\cal L}} |leaves(\tilde T)|$ where\\[1ex]
${\cal L}=\{\tilde T \mid \tilde T  \text{ is directed spanning tree for } G'_\BN \text{with root } r'\}
$
and $G'_\BN=(V(G') \cup \{r',y\},A(G') \cup\left(\{(r',y)\} \cup_{u \in
V_\BN} (r',u) \right)$ and $r',y$ are new vertices.\\
 Entries where such a directed spanning tree $\tilde T$ does not exits (e.g. if $V_\BN=\emptyset$) get the value $\emptyset$.
This table can be filled up in time $\Oh^*({n \choose \alpha \cdot n}\cdot 2^{\alpha n} \cdot \rtdmlst^{\alpha n})\subseteq \Oh^*(1.8139^n)$. This
running time is composed of enumerating $SG^\alpha$, then by cycling through all possibilities for $V_\BN$ and finally solving the problem on
instance $G'_\BN$ with
Algorithm~\ref{dalgo1}. \\
\begin{theorem}
 {\sc Directed Maximum Leaf Spanning Tree} can be solved in time $\Oh^*(1.8139^n)$ consuming $\Oh^*(1.6563^n)$ space.
\end{theorem}
\begin{proof}
Run the above mentioned $\Oh^*(2^n)$-algorithm until $|G^r| \le \alpha \cdot n$ with $G^r:=V \setminus internal(T)$. Then let $T^e=L[G^r,V(G^r) \cap
\BN_T]$. Note that the vertex $r \in V(T^e)$ must be internal and $y \in leaves(T^e)$. By Lemma~\ref{drosslem1} we can assume that
$A(\{r\} \cup \nout(r)) \subseteq T^e$.
Now identify the vertices $BN_T \cap V(T^e)$ with $V(G^r) \cap \BN_T$ and delete $r$ an $y$ to a directed spanning 
tree $\hat T$ for the original graph $G$. Or more formally let
$\hat
T:=T \cup (T^e \setminus A(\{r\} \cup \nout(r))$. Observe that $\hat T$ extends $T$ to optimality.
\qed
 \end{proof}

Note that in the first phase we cannot substitute the  $\Oh^*(2^n)$-algorithm by Algorithm~\ref{dalgo1}.
 It might be the case that {\bf (R6)}
generates graphs which are not vertex-induced subgraphs of $G$.
\subsection{R\'esum\'e}
The paper at hand presented an algorithm which solves the {\sc Directed Maximum Leaf Spanning Tree} problem in time $\Oh^*(\rtdmlst^n)$. Although this
algorithm follows the same line of attack as the one of \cite{Feretal2009b} the algorithm itself differs notably. 
The approach of
\cite{Feretal2009b} does not simply carry over. To achieve our run time bound we had to develop new algorithmic ideas. This is reflected by the
greater number of branching cases. 
\bibliographystyle{plain}


\begin{thebibliography}{10}

\bibitem{Bluetal2005}
J.~Blum, M.~Ding, A.~Thaeler, and X.~Cheng.
\newblock Connected dominating set in sensor networks and MANETs,
 \newblock {\em Handbook of Combinatorial Optimization, Vol.~B}, pp. 329--369. Springer, \longversion{Heidelberg, }2005.




\bibitem{BonZic2008}
P.~S. Bonsma and F.~Zickfeld.
\newblock A 3/2-approximation algorithm for finding spanning trees with many
  leaves in cubic graphs.
\newblock In {\em WG}, 
{\em LNCS} 5344:
66--77. Springer, \longversion{Heidelberg, }2008.




\bibitem{DalGutKimYeo2008b}
J.~Daligault, G.~Gutin, E.~J. Kim, and A.~Yeo.
\newblock {FPT} algorithms and kernels for the directed $k$-leaf problem.
\newblock In {\em Journal of Computer and System Sciences}, 2009.
\newblock http://dx.doi.org/10.1016/j.jcss.2009.06.005


\bibitem{DalTho2009}
J.~Daligault and S.~Thomass\'e.
\newblock On finding directed trees with many leaves.
\newblock In {\em IWPEC}, 2009, to appear.





\bibitem{Feretal2009b}
H. Fernau, A. Langer, M. Liedloff, J. Kneis, D. Kratsch, D. Raible and P. Rossmanith.
\newblock An exact algorithm for the Maximum Leaf Spanning Tree problem.
\newblock In {\em IWPEC}, 2009, to appear.




\bibitem{FomGraKra08}
F.V. Fomin, F.~Grandoni and D.~Kratsch.
\newblock Solving Connected Dominating Set Faster than 2$^{\mbox{{\it n}}}$.
\newblock {\em Algorithmica}, 52(2):153--166, 2008.

\bibitem{FomGraKra2009}
F.V.~Fomin, F.~Grandoni and D.~Kratsch.
\newblock A measure {\&} conquer approach for the analysis of exact algorithms.
\newblock   {\em Journal of the  ACM}, 56(5),2009.


\bibitem{KneLanRos2008a}
J.~Kneis, A.~Langer, and P.~Rossmanith.
\newblock A new algorithm for finding trees with many leaves.
\newblock In {\em ISAAC}, 
{\em LNCS} 5369:
270--281. Springer, \longversion{Heidelberg, }2008.

\bibitem{KouWil2009}
I.~Koutis and R.~Williams.
\newblock Limits and Applications of Group Algebras for Parameterized Problems.
\newblock In {\em ICALP (1)}, {\em LNCS} 5555:653--664. Springer, \longversion{Heidelberg, }2009.
 

\bibitem{LuRav98}
H.-I Lu and R.~Ravi.
\newblock Approximating maximum leaf spanning trees in almost linear time.
\newblock {\em Journal of Algorithms} 29:132--141, 1998.

\bibitem{RaiFer2010}
D.~Raible and H.~Fernau.
\newblock An Amortized Search Tree Analysis for $k$-Leaf Spanning Tree
\newblock In {\em  International Conference on Current Trends in Theory and Practice of Computer Science  (SOFSEM)}, 2010, to appear.
\bibitem{RamSau2006}
V. Raman and S. Saurabh.
\newblock Parameterized algorithms for feedback set problems and their duals in tournaments,
\newblock In {\em Theoretical Computer Science}, 351(3):446--458, 2006.

\bibitem{Sol-Oba98}
R.~Solis-Oba.
\newblock 2-approximation algorithm for finding a spanning tree with maximum
  number of leaves.
\newblock In {\em ESA}, 
{\em LNCS} 1461:
441--452. Springer, \longversion{Heidelberg, }1998.

\bibitem{Thaetal2007}
M.~T. Thai, F.~Wang, D.~Liu, S.~Zhu, and D.-Z. Du.
\newblock Connected dominating sets in wireless networks different transmission
  ranges.
\newblock {\em IEEE Trans. Mobile Computing} 6:1--10, 2007.

\end{thebibliography}

\end{document}